# Amorphous selenium-alloyed $TeO_x$ thin-film solar cells

Zhaoheng Ling[1,2,*†], Linqu Luo[1,2,*†], Chrysa Aivalioti[2], Temur Maksudov[2], Mengxi Zhang[1], Theodoulos Kalogerakis[1], Lazaros Panagiotidis[2], Elefterios Lidorikis[3], Spyros Doukas[3], Nikolaos Kalfagiannis[3], Epameinondas Tziortzios[4,5], Leonidas Tsetseris[5], Huihui Zhu[6], Ao Liu[7], Yong-Young Noh[8], Martin Heeney[2], Shadi Fatayer[2], and Thomas D. Anthopoulos[1]*

[1] Henry Royce Institute and Photon Science Institute, Department of Electrical and Electronic Engineering, The University of Manchester, Oxford Road, Manchester M13 9PL, United Kingdom.

[2] Physical Science and Engineering Division, King Abdullah University of Science and Technology (KAUST), Thuwal 23955-6900, Saudi Arabia.

[3]Department of Materials Science and Engineering, University of Ioannina, Ioannina 45110, Greece.

[4]University of Rennes, CNRS, Institute des Sciences Chimiques de Rennes, UMR 6226, Rennes, France.

[5]Department of Physics, School of Applied Mathematical and Physical Sciences, National Technical University of Athens, Athens GR-15780, Greece

[6]Department of Physics, University of Electronic Science and Technology of China, Chengdu, China.

[7]Institute of Fundamental and Frontier Sciences, State Key Laboratory of Electronic Thin Films and Integrated Devices, University of Electronic Science and Technology of China, Chengdu, China.

[8]Department of Chemical Engineering, Pohang University of Science and Technology, Pohang, Gyeongbuk, Republic of Korea.

* **Correspondence**:
Zhaoheng Ling (zhaoheng.ling@manchester.ac.uk)
Linqu Luo (linqu.luo@manchester.ac.uk)
Thomas D. Anthopoulos (thomas.anthopoulos@manchester.ac.uk)
† Zhaoheng Ling and Linqu Luo contributed equally to this work.

**Abstract**

Tellurium dioxide ($TeO_2$) is a wide-bandgap oxide with attractive processing versatility and emerging p-type semiconducting behaviour, making it a promising material system for optoelectronic and photovoltaic applications. However, $TeO_2$-derived thin films prepared by scalable manufacturing processes, such as vacuum sublimation, can contain residual metallic $Te^0$ species that introduce trap states, limit hole transport, and promote carrier recombination. Here, we demonstrate a selenium alloying strategy to control the defect chemistry of vacuum-sublimed $TeO_x$ films. $TeO_2$:Se powder mixtures (90:10 wt%) were thermally evaporated and subsequently annealed in air at 225 °C for up to 9 h, yielding amorphous $TeO_x$:Se films with an optical bandgap of 0.98 eV under optimised conditions. Structural, chemical, optical, and electrical analyses show that prolonged annealing promotes Se redistribution, suppresses residual $Te^0$ to approximately 0.4%, reduces the film's surface roughness and shifts the valence-band edge. To evaluate the potential of $TeO_x$:Se for photovoltaic applications, we fabricated thin-film solar cells with an ITO/ZnO/$TeO_x$:Se/$MoO_x$/Ag architecture. Devices based on an 80 nm $TeO_x$:Se active layer annealed for 9 h exhibit optimal performance, with broadband photocurrent generation from 300 to 1000 nm, yielding an open-circuit voltage of 0.50 V, a short-circuit current density of 8.50 mA $cm^{-2}$, a fill factor of 0.44, and a maximum power conversion efficiency of circa 1.87%. These results establish selenium-alloyed $TeO_x$ as a viable oxide-derived absorber for thin-film photovoltaics and highlight chemical-state control as an effective route for improving the photovoltaic performance.

## Introduction

Thin-film photovoltaic technologies such as cadmium telluride and copper indium gallium selenide have achieved high efficiencies through continuous advances in absorber design, interface engineering, and device architecture.[1–8] At the same time, expanding the absorber materials portfolio beyond conventional thin-film semiconductors remains important, particularly because emerging systems such as organic photovoltaics still face practical constraints related to morphology sensitivity, narrow processing windows, and operational stability under heat, oxygen, and illumination.[9–12] These challenges motivate the exploration of alternative active-layer concepts that can reduce reliance on delicate bulk-heterojunction morphology while retaining manufacturable processing routes such as vacuum deposition and low-temperature thermal treatment. In this context, an oxide-derived absorber that can be deposited reproducibly and whose energetics can be tuned by chemical composition and thermal annealing offers a complementary pathway for inverted photovoltaic devices.

Tellurium dioxide ($TeO_2$) is an appealing material for vacuum-processed photovoltaic absorbers because of its chemical robustness and compatibility with thermal evaporation. Controlled reduction and bonding reconstruction can convert $TeO_2$-derived films into semiconducting $TeO_x$ networks, in which Te-rich bonding motifs introduce Te 5p-derived states near the valence-band edge that support hole transport.[13–15] Recent work has shown that $TeO_2$-derived films can be converted into semiconducting Te suboxide networks through controlled reduction and bonding reconstruction at modest temperatures.[16] For photovoltaic devices, such a $TeO_2$ platform is attractive because: (i) vacuum sublimation offers scalable processing of uniform films and straightforward thickness control; (ii) oxide chemistry can improve tolerance to ambient exposure relative to other absorber families such as organics and perovskites; and (iii) the electronic structure is highly sensitive to local Te-oxygen versus Te-Te bonding motifs, allowing for versatile energetic tuning.[17] However, the same chemical flexibility imposes an intrinsic limitation: Te exhibits multiple oxidation states, where incomplete oxidation or local phase separation can yield reduced metallic $Te^0$ species and oxygen-deficiency-related sub-gap states.[18,19] Although in conventional electronic devices, such as transistors, this mixed-valence character can be used to improve charge transport, in photovoltaic cells it can be detrimental, because $Te^0$-related states can pin the Fermi energy, increase trap-assisted recombination and introduce shunting pathways that limit the device's open-circuit voltage ($V_{oc}$) and fill factor (FF).[20] Therefore, a central requirement for $TeO_2$ active layers is to stabilise the oxidation-state distribution and suppress the formation of metallic Te0 without sacrificing optical absorption and carrier extraction.

Selenium (Se) provides a chemically and electronically compatible route to address this challenge. As an isovalent chalcogen, Se can alloy with Te to reshape local bonding preferences and tune band structure relevant to solar cell applications.[20] In $TeO_2$-derived semiconductors, Se alloying has also been shown to regulate carrier concentration and suppress trap-limited transport, enabling high-performance devices in related platforms.[21–23] Building on this knowledge, we hypothesised that Se incorporation into $TeO_2$ can suppress the formation of metallic $Te^0$ and associated sub-gap defect states, while thermal annealing can drive Se redistribution within the Te-O amorphous network, thereby improving film uniformity and its energetics.[24]

Herein, we deposited $TeO_x$:Se layers and evaluated different parameters, including Se concentration, film thickness, and annealing conditions. Using complementary chemical-state and optoelectronic characterisation techniques spanning depth-sensitive bonding analysis, photoelectron spectroscopy (PESA)-derived valence-band maximum (VBM) measurements, optical absorption/Tauc analysis, and recombination/transport diagnostics, we find that optimised annealing suppresses $Te^0$, modulates the band-edge energetics, and reduces trap-assisted recombination.[25] Photovoltaic devices featuring an 80 nm-thick $TeO_x$:Se (90:10%) absorber layer annealed for 9 h achieve the best power conversion efficiency (PCE) of 1.87% in an ITO/ZnO/TeOx (TeOx:Se)/MoOx/Ag device architecture. Our work highlights the Se-mediated defect-mitigation strategy for $TeO_2$ as a promising route to efficient photovoltaics.

## Results and discussion

We followed a thermal evaporation and air-annealing workflow to prepare all $TeO_x$ and $TeO_x$:Se active layers with controlled thickness and post-deposition treatment, as illustrated in **Fig. 1a**. This approach intentionally separates deposition from chemical equilibration: the evaporation step yields continuous $TeO_x$-derived films, while annealing provides a defined kinetic window for atomic rearrangement, oxidation-state equilibration, and Se redistribution within the amorphous network. We used $TeO_x$ films as a reference for TeOx:Se(10%) layers, where 10% denotes a nominal 10 wt% Se in the powder mixture. As-prepared TeOx:Se layers were then annealed for 1.5 h and 9 h, representing insufficient and optimal annealing conditions, respectively. This simple methodology enabled correlation among film processing, chemical states, energetics, and device operation.

To evaluate the morphology of the $TeO_2$ and $TeO_x$:Se layers before and after thermal treatment, we performed optical microscopy (OM), atomic force microscopy (AFM), and scanning electron microscopy (SEM). Representative AFM/OM and SEM images are presented **in Fig. 1b**, with the surface height histogram shown in **Fig. 1c** and additional morphology sets provided in **Figs. S1-S3**. The $TeO_x$ reference film exhibits a surface with RMS roughness of 7.36 nm (**Fig. 1b**). After introducing 10 wt% Se in the $TeO_2$:Se powder mixture and applying a short 1.5 h post-deposition annealing step, the surface becomes visibly more corrugated and heterogeneous, with RMS roughness increasing to circa 11.05 nm. Extending the annealing time to 9 h yields films with the smoothest surface (RMS roughness of ∼6 nm), as evidenced by the narrower surface height distribution in **Fig. 1c**. In addition to the nanoscopic topographic changes, the OM images show a clear variation in the macroscopic optical contrast after Se incorporation and annealing. The more uniform colour tone observed in the 9 h film implies a more homogeneous layer across the substrate, which is consistent with its lower roughness and denser morphology revealed by AFM.

SEM imaging supports the same trend: the $TeO_2$ reference shows a characteristic pinhole-free morphology, while $TeO_x$:Se films exhibit annealing-dependent densification and grain/cluster evolution.[22,26–28] The 1.5 h annealed layer retains more pronounced features and less uniform packing, whereas the 9 h film exhibits a more compact, homogeneous microstructure, consistent with improved film continuity and fewer void-like regions. From these measurements, we conclude that the 9 h annealed film displays a denser, more continuous surface micro-texture than the 1.5 h film (**Fig. 1b**, bottom row). The series of SEM images in **Figs. S1-S3** further confirm this conclusion.

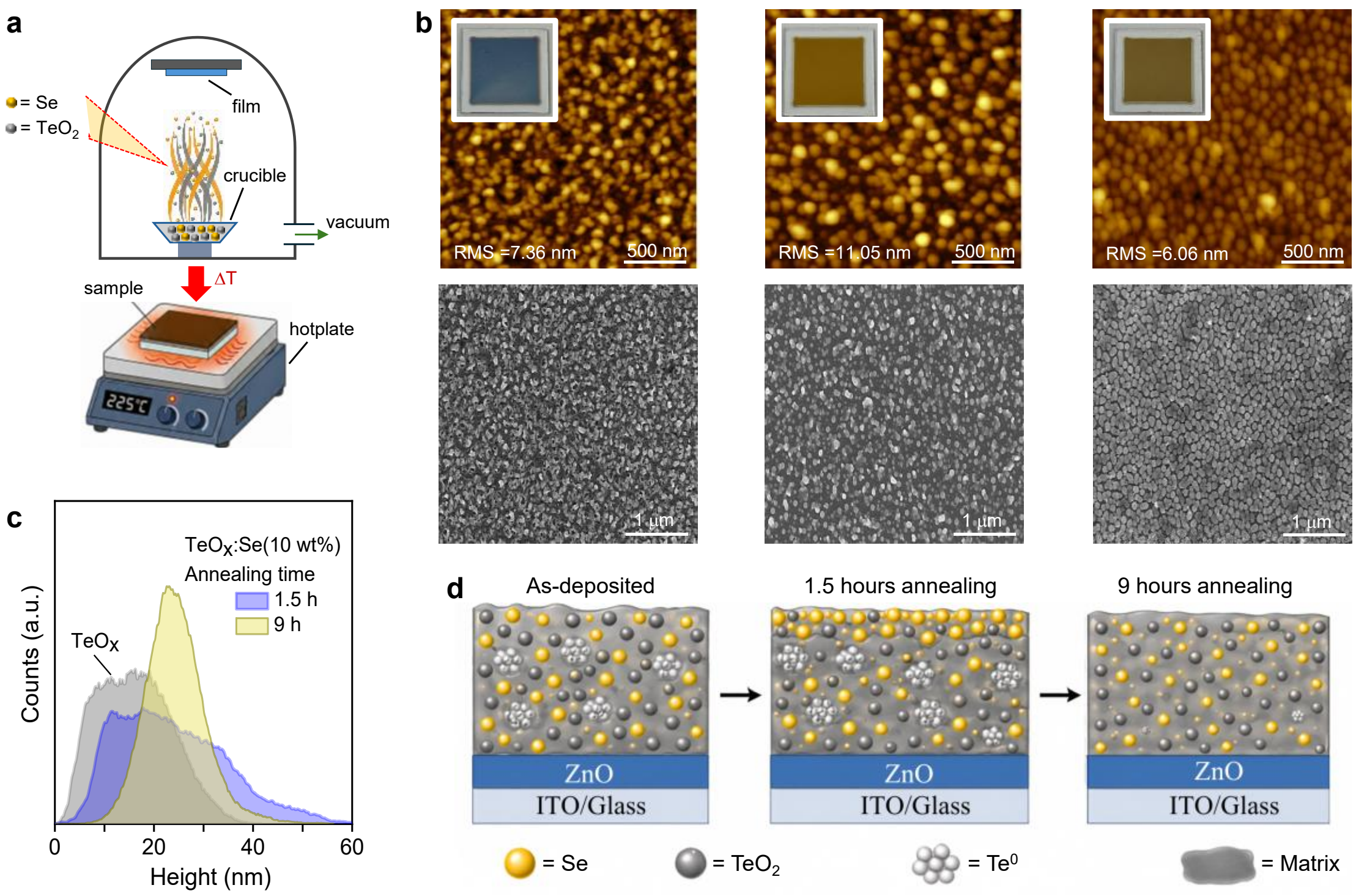

**Figure 1**. (a) Schematic diagram of thin film thermal evaporation and annealing, (b) AFM topography images (top row) of $TeO_x$, $TeO_x$:Se(10 wt% Se) annealed for 1.5 h, and $TeO_x$:Se(10 wt% Se) annealed for 9 h, arranged from left to right. Insets in the AFM images are the photographs of the corresponding samples. SEM images of the same films are shown in the bottom row in the same sequence. (c) Surface height histograms of different films. (d) Schematic illustration of annealing-driven Se redistribution and $Te^0$ suppression in $TeO_x$:Se films.

As discussed above, Te can adopt multiple oxidation states, and residual metallic $Te^0$ species can introduce recombination-active electronic states that compromise photovoltaic operation. We therefore proposed that prolonged annealing in air would promote Se redistribution and chemical integration within the amorphous Te-O network, while concurrently suppressing metallic $Te^0$. This mechanistic picture, schematically illustrated in **Fig. 1d**, predicts that insufficiently annealed layers should retain a higher $Te^0$ content, whereas extended annealing should reduce the concentration of residual metallic $Te^0$. Such chemical-state control is expected to modify the material energetics, including the optical bandgap, band-edge positions, and Fermi level, while reducing trap-assisted carrier recombination. Together, these effects should lead to devices with improved operating characteristics.

To test these predictions, angle-resolved X-ray photoelectron spectroscopy (XPS) was performed on the $TeO_x$-derived films to quantify the chemical states of Te and evaluate the angle-dependent elemental composition. Deconvolution of the Te 3d region was used to extract the metallic $Te^0$ fraction at multiple take-off angles; the core-level spectra are shown in **Figs. S5-S6**, and the extracted percentages are summarised in **Tables S1-S3**. Across all films, oxidised Te components dominate the Te 3d envelope, consistent with a $Te^{4+}$-rich oxide framework formed under air annealing. This is important because it confirms that selenium incorporation does not simply transform the active layer into a metallic Te film; instead, the films remain oxide-dominated while containing varying amounts of a reduced $Te^0$ component.[22,29] The $TeO_x$ reference shows residual $Te^0$ even in the nominal oxide, with $Te^0$ fractions of 0.8%, 0.7%, and 0.6% at take-off angles 90°, 120°, and 150° (**Table S1**), implying incomplete oxidation or local reduction during deposition/annealing that could seed defect states at interfaces. After introducing 10 wt% Se and annealing for 1.5 h, $Te^0$ increases substantially to 1.7%, 1.6%, and 1.4% at 90°, 120°, and 150° (**Table S2**), indicating kinetically limited oxidation and chemical heterogeneity under insufficient annealing. By contrast, after 9 h of annealing, $Te^0$ is strongly suppressed to approximately 0.4% at all three take-off angles (**Table S3**). The suppression is consistent across angles, indicating that the reduction of $Te^0$ is not a surface effect but reflects a broadly improved chemical configuration in the near-surface region. From a device perspective, reducing the amount of $Te^0$ is expected to boost $V_{oc}$ by lowering trap-assisted recombination at the interface with the carrier-extracting contact and to raise FF by reducing defect-assisted recombination and carrier transport improvements. The fact that $Te^0$ decreases from 1.6% to 0.4% (120°) and from 1.4% to 0.4% (150°) provides quantitative support for a substantial decrease in the reduced $Te^0$ phase that could otherwise dominate interfacial carrier dynamics.

To further assess the chemical environment of Se, we examined the take-off-angle dependence of Se signals extracted from the same datasets (**Tables S2-S3**). In the 1.5 h film, the Se atomic fraction varies more strongly with angle (5.2%, 4.9%, and 5.7% at 90°, 120°, and 150°), suggesting a measurable angle-dependent Se variation within the XPS probing depth, consistent with incomplete Se homogenization after short annealing. In the 9 h film, the Se fraction remains angle-dependent but shows a less monotonic profile (4.4%, 4.4%, and 5.1%), consistent with a more equilibrated redistribution state after longer annealing. While the angle-resolved geometry does not, by itself, provide an absolute depth profile, these results, together with observed $Te^0$ suppression, suggest annealing-modulated Se redistribution and chemical integration within the Te-O network.

The X-ray diffraction (XRD) patterns of the different films were also recorded and shown in **Fig. S7**. The diffractogram is consistent with largely amorphous or weakly ordered films, and no pronounced new crystalline signatures emerge that would alone account for the previously observed energetic shifts and device improvements.[22] This observation strengthens the interpretation that local bonding and defect tuning, rather than long-range crystallinity-induced changes, primarily affect the energy levels in $TeO_x$:Se layers.

The optical and electronic structures of the $TeO_x$-based absorbers were subsequently examined to establish how Se incorporation and thermal annealing regulate the material energetics. **Fig. 2a** compares the normalised UV-vis-NIR absorption spectra of representative pristine $TeO_x$ and $TeO_x$:Se(90:10 wt%) films. Se incorporation results in pronounced changes in the absorption spectral with optical absorption extending well into the near-infrared region. Prolonging the annealing time from 1.5 to 9 h further shifts the principal absorption feature toward shorter wavelengths and redistributes the spectral response across the visible-to-near-infrared region, highlighting the major impact of thermal annealing on the optical properties of the $TeO_x$:Se layer. The complete absorption data as a function of film thickness, annealing duration, and Se concentration are shown in **Fig. S8,** while **Fig. 2b** display direct bandgap Tauc plot analysis for the same absorption spectra. Extrapolating the linear part of the plot to the photon energy axis yields optical bandgaps of 0.86 eV for $TeO_x$, 1.09 eV for $TeO_x$:Se(10 wt%) annealed for 1.5 h, and 0.98 eV for $TeO_x$:Se(10 wt%) annealed for 9 h. Evidently, Se incorporation widens the optical gap relative to pristine $TeO_2$, whereas prolonged annealing induces a modest narrowing within the Se-alloyed films.

The VBM positions were independently determined by PESA, with the corresponding photoemission-yield plots presented in **Fig. S4** and the extracted values summarised in **Table S4**. The VBM values, with respect to the vacuum level, are -5.20 eV for $TeO_x$, -5.54 eV for $TeO_x$:Se(10 wt%) annealed for 1.5 h, and -5.45 eV for $TeO_x$:Se(10 wt%) annealed for 9 h. Combining these PESA-derived VBM values with the experimentally determined optical bandgaps gives estimated CBM positions of -4.34, -4.45, and -4.47 eV, respectively (**Fig. 2c**). Notably, prolonged annealing shifts the VBM of $TeO_x$:Se(10 wt%) from -5.54 to -5.45 eV, whereas the estimated CBM remains essentially unchanged within the experimental and fitting uncertainty. The accompanying reduction in the optical bandgap from 1.09 to 0.98 eV therefore indicates that the annealing-induced evolution of the band-edge energetics is primarily associated with changes in the valence band rather than with a substantial shift in the conduction band edge.

The evolution in optical and electronic energetics occurs concurrently with the pronounced chemical-state changes identified by XPS. In particular, prolonged annealing reduces the residual $Te^0$ fraction from approximately 1.4-1.7% to approx. 0.4% across the three measurement geometries, while simultaneously modifying the Se distribution vertically. The coincidence of $Te^0$ suppression, Se redistribution, VBM displacement, and optical-gap narrowing indicates that thermal treatment reorganises the local chemical environment and associated electronic states within the amorphous $TeO_x$:Se network. To gain further insight into the role of Se incorporation, density functional theory (DFT) calculations were performed using oxygen-deficient $\alpha$-$TeO_2$ supercells as atomistic models. The calculations indicate that oxygen vacancies tend to aggregate into Te-rich local complexes, while Se atoms preferentially occupy oxygen-deficient regions and form Se-Te bonding configurations. These structural modifications are accompanied by changes in the calculated electronic bandgap, highlighting the sensitivity of the local bonding environment and electronic structure to Se incorporation (see Supporting Information for computational details). Collectively, these results establish a correlation between annealing-induced chemical-state regulation and the evolution of the optical and electronic structure of $TeO_x$:Se.

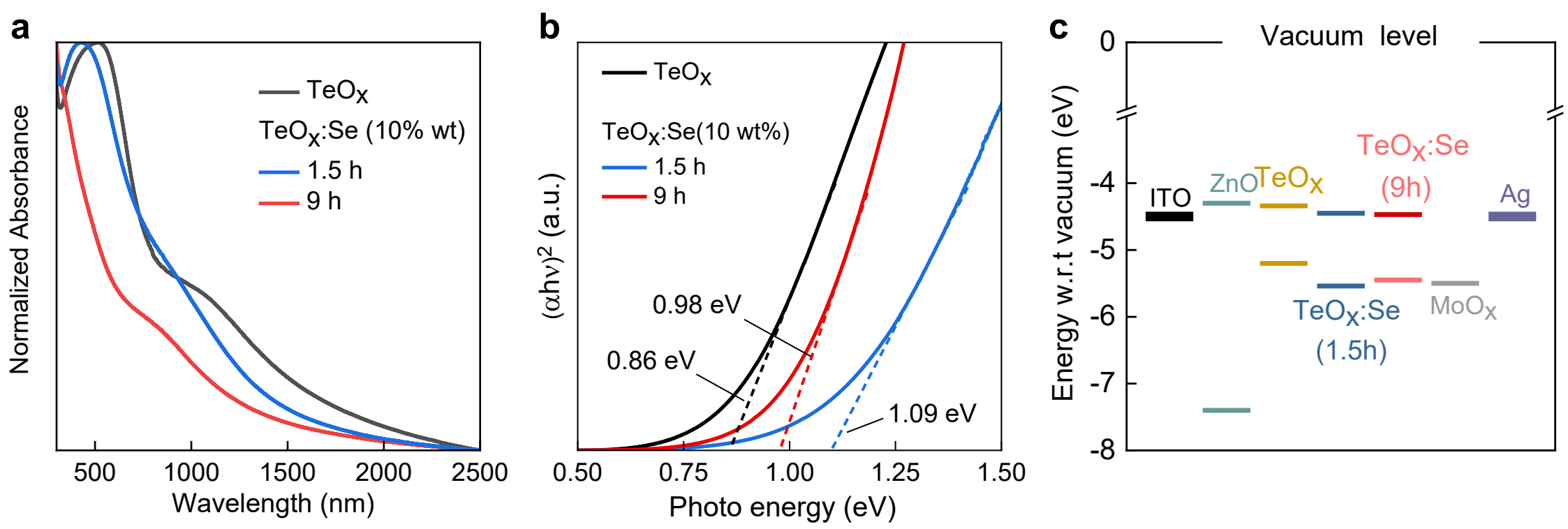

**Figure 2**. (a) Normalised UV–vis–NIR absorption spectra of $TeO_x$ annealed for 9 h and $TeO_x$:Se(10 wt%) films annealed for 1.5 and 9 h. (b) Corresponding direct Tauc plots and extracted optical bandgap energies. (c) Energy-level diagram constructed from PESA-derived VBM positions and Tauc-derived optical bandgaps. The energy levels of materials relevant to solar cell fabrication (i.e., ITO, ZnO, MoOx, and Ag) are also shown in (c) for comparison.

To evaluate the potential of $TeO_x$:Se as a photovoltaic absorber, we fabricated solar cells with an ITO/ZnO/$TeO_x$:Se/$MoO_x$/Ag architecture (**Fig. 3a**). We systematically optimised the active-layer thickness, annealing time, and Se concentration. The *J–V* characteristics and corresponding photovoltaic parameters are presented in **Figs. 3b-d** and **Tables S5-S7**. Active layer thickness experiments identified 80 nm as the optimal thickness for the $TeO_x$:Se(10 wt%) layer within the explored range (**Fig. 3b**; **Table S5**), offering a balance between sufficient optical absorption, which enhances the short-circuit current ($J_{SC}$), and improved charge transport, which increases the fill factor (FF) and reduces shunt-resistance-related losses. Cells annealed for 1.5 h yielded $V_{OC}$ = 0.26 V, $J_{SC}$ = 5.69 mA cm$^{-2}$, FF = 0.35, and PCE = 0.52%, whereas the device annealed for 9 h exhibited $V_{OC}$ = 0.50 V, $J_{SC}$ = 8.50 mA cm$^{-2}$, FF = 0.44, and PCE = 1.87% (**Fig. 3c**, **Table S6**). The simultaneous increases in $V_{OC}$, $J_{SC}$, and FF are consistent with reduced recombination losses and improved carrier collection. Composition optimisation further highlights selenium's enabling role. Specifically, among the tested Se concentrations under the optimised annealing protocol (**Fig. 3d**, **Table S7**), 10 wt% Se provided the best results, indicating that sufficient Se incorporation is required for defect suppression and carrier transport optimisation.

External quantum efficiency (EQE) measurements were performed to verify that the photocurrent enhancement arises from improvements in photogeneration and carrier collection (**Fig. 3e**). The optimally annealed cell (9 h) exhibits consistently higher EQE than the device annealed for 1.5 h, and the integrated current agrees with the $J_{SC}$ extracted from the *J–V* characteristics under simulated solar irradiation. This agreement confirms that the increase in $J_{SC}$ from 5.69 to 8.50 mA cm$^{-2}$ reflects an enhanced photoresponse and improved carrier extraction. These concurrent improvements are consistent with the morphology and chemical-state analyses presented earlier. Specifically, the smoother film surface (RMS roughness of ~6 nm) and the lower residual $Te^0$ (0.4%) reduced recombination losses during carrier photogeneration and transport, thereby increasing the fraction of photogenerated carriers collected under short-circuit conditions.

To directly probe recombination kinetics and determine whether trap-assisted recombination is mitigated by annealing-driven Se redistribution, we performed transient photovoltage (TPV) measurements (**Fig. 3f**) and light-intensity-dependent analyses (**Fig. S9**). The TPV lifetime increased from ~3.7 μs for the 1.5 h annealed device to ~5.5 μs for the 9 h annealed device, indicating slower recombination kinetics with increasing annealing time.[30] Light-intensity-dependent measurements provide further mechanistic insight. The $J_{SC}$–P slope increased from 0.88 to 0.94, approaching ideal linearity and indicating improved carrier collection, whereas the $V_{OC}$–ln(I) slope decreased from 2.20 to 1.72 kT/q, consistent with a reduced contribution from trap-assisted recombination.[31] Thus, the spectroscopically observed defect suppression correlates with improved optoelectronic characteristics, including an increase in $V_{OC}$ from 0.26 to 0.50 V and an increase in FF from 0.35 to 0.44 after 9 h of thermal annealing.

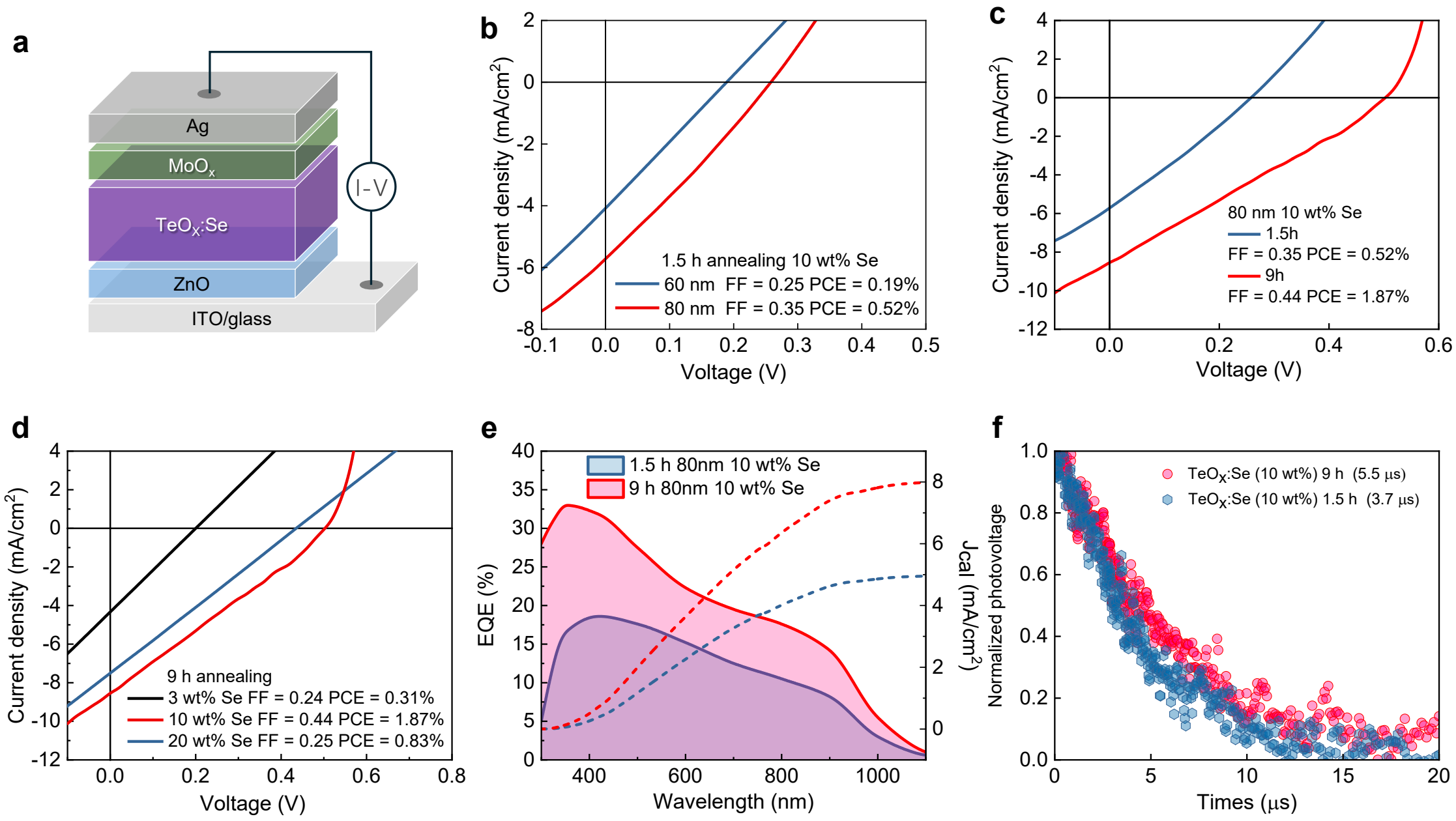

**Figure 3**. (a) Schematic structure of the solar cells developed in this study. *J–V* characteristics of devices based on (b) $TeO_x$:Se(90:10 wt%) films with different thicknesses annealed for 1.5 h, (c) 80 nm $TeO_x$:Se(90:10 wt%) films annealed for different times, and (d) $TeO_x$:Se films with different Se concentrations annealed for 9 h. (e) EQE spectra of devices based on 80 nm $TeO_x$:Se(10 wt% Se) films annealed for different times. (f) Transient photovoltage (TPV) spectra of devices based on 80 nm $TeO_x$:Se(90:10 wt%) films annealed for 1.5 and 9 h.

Electrical impedance measurements were obtained for photovoltaic cells prepared with varying absorber thickness, annealing time, and Se concentration (**Fig. S10**) in order to probe charge transport and interfacial losses.[32] The cell with the optimal response exhibits a more favourable impedance profile, consistent with improved charge transport and reduced interfacial recombination. This behaviour agrees with the transient and light-intensity-dependent analysis, which independently indicate slower recombination kinetics, enhanced carrier collection, and reduced trap-assisted recombination. Such transport improvements are directly linked to the higher FF through improved and balanced carrier extraction, and they may also contribute to the higher $V_{OC}$ by suppressing interfacial recombination losses. Together, these results reinforce the close interdependence of film morphology, chemical state, energetics, charge transport, and photovoltaic device operation.

## Conclusion

This study demonstrates selenium-enabled defect control as an effective strategy for improving amorphous $TeO_x$-derived photovoltaic absorbers. Thermal annealing promotes Se redistribution within the Te-O network, suppresses residual metallic $Te^0$, improves film morphology, shifts the valence-band edge from -5.54 to -5.45 eV, and narrows the optical bandgap from 1.09 to 0.98 eV. These changes indicate that annealing reorganises the local bonding environment and reduces defect-mediated electronic disorder. Devices based on an 80 nm $TeO_x$:Se(90:10 wt%) absorber annealed for 9 h achieved a maximum PCE of 1.87%, with $V_{OC}$ = 0.50 V, $J_{SC}$ = 8.50 mA $cm^{-2}$, and FF = 0.44, representing an approximately fourfold improvement over the 1.5 h control cell. TPV and light-intensity measurements confirm slower recombination and improved carrier collection for optimally annealed cells. Overall, our findings establish $TeO_x$:Se as a promising oxide-derived absorber and highlight chemical-state engineering as a route towards improved thin-film photovoltaics.

## Authorship contribution statement

Z.L., L.L., and T.D.A. conceived and conceptualised the project. Z.L. and L.L. designed the experimental framework and coordinated the overall research progress. Z.L. and L.L. fabricated the devices. Z.L., M.Z, and T.K. performed the solar cell characterisations and analysed the results. C.A. and S.F. performed and analysed the XPS measurements. T.M. and L.L. conducted XRD measurements and analysed the results. L.L. carried out AFM topography measurements. L.T. and E.T performed the DFT calculation. E.L., S.D., N.K., H.Z., A.L., Y.N. and M.H. contributed to critical discussions and manuscript preparation. Z.L., L.L. and T.D.A. designed and prepared the schematics, and Z.L. wrote the first draft of the manuscript. T.D.A. supervised the project. All authors contributed to data interpretation, critically revised the manuscript, and approved the final version.

**Acknowledgements**

We acknowledge the support of the Henry Royce Institute established through EPSRC grants EP/R00661X/1, EP/P025498/1, EP/P025021/1, and EP/S019367/1, and the National Research Foundation of Korea (NRF) grant funded by the Korea government (MSIT) (RS-2023-00260608). TDA acknowledge financial support from UK Research and Innovation (UKRI) and the European Research Council (ERC), grant reference number EP/Z534006/1.

**Amorphous selenium-alloyed $TeO_x$ thin-film solar cells**

Zhaoheng Ling[1,2,*†], Linqu Luo[1,2,*†], Chrysa Aivalioti[2], Temur Maksudov[2], Mengxi Zhang[1], Theodoulos Kalogerakis[1], Lazaros Panagiotidis[2], Elefterios Lidorikis[3], Spyros Doukas[3], Nikolaos Kalfagiannis[3], Epameinondas Tziortzios[4,5], Leonidas Tsetseris[5], Huihui Zhu[6], Ao Liu[7], Yong-Young Noh[8], Martin Heeney[2], Shadi Fatayer[2], and Thomas D. Anthopoulos[1]*

[1] Henry Royce Institute and Photon Science Institute, Department of Electrical and Electronic Engineering, The University of Manchester, Oxford Road, Manchester M13 9PL, United Kingdom.

[2] Physical Science and Engineering Division, King Abdullah University of Science and Technology (KAUST), Thuwal 23955-6900, Saudi Arabia.

[3]Department of Materials Science and Engineering, University of Ioannina, Ioannina 45110, Greece.

[4]University of Rennes, CNRS, Institute des Sciences Chimiques de Rennes, UMR 6226, Rennes, France.

[5]Department of Physics, School of Applied Mathematical and Physical Sciences, National Technical University of Athens, Athens GR-15780, Greece

[6]Department of Physics, University of Electronic Science and Technology of China, Chengdu, China.

[7]Institute of Fundamental and Frontier Sciences, State Key Laboratory of Electronic Thin Films and Integrated Devices, University of Electronic Science and Technology of China, Chengdu, China.

[8]Department of Chemical Engineering, Pohang University of Science and Technology, Pohang, Gyeongbuk, Republic of Korea.

* **Correspondence**:

Zhaoheng Ling (zhaoheng.ling@manchester.ac.uk)

Linqu Luo (linqu.luo@manchester.ac.uk)

Thomas D. Anthopoulos (thomas.anthopoulos@manchester.ac.uk)

† Zhaoheng Ling and Linqu Luo contributed equally to this work.

**Films depositions and photovoltaic fabrication:** Glass substrates with prepatterned ITO were first cleaned by sequential ultrasonication in a dilute Extran 300 detergent solution and deionized water for 30 min, respectively. The substrates were cleaned with acetone and isopropanol by ultrasonication for 10 min each. Next, the substrates were exposed to ultraviolet-ozone treatment for 20 min. Zinc acetate dihydrate (100 mg) was dissolved in 2-methoxyethanol (4 mL) with ethanolamine (28 μL) as a stabilizer. The zinc acetate dihydrate solution was stirred at 60 °C for at least 6 h. The ZnO thin films were deposited by spin-coating a zinc acetate solution at 6000 rpm for 30 s. The films were thermally annealed at 200 °C for 20 min. $TeO_2$ (≥97%) and Se (99.99%) powders were purchased from Sigma-Aldrich and used as received. Prior to evaporation, $TeO_2$ and Se powders were mixed at the desired weight ratios and ground for 5 minutes. For instance, to achieve a Se content of 10 wt%, 40 mg of Se and 360 mg of $TeO_2$ were mixed and loaded into a tungsten boat for thermal evaporation. The evaporation was integrated inside $N_2$-filled glovebox to minimize the possible contaminations and mitigate vapour toxicity. The substrate temperature was maintained at room temperature, and the base pressure prior to deposition was around $2 \times 10^{-6}$ Torr. The films thicknesses were monitored during deposition and the shutter was closed once the desired thickness was reached. The deposited $TeO_x$:Se films were subsequently annealed at 225 °C for the desired duration time in ambient air. Finally, the samples were transferred into a thermal evaporator for the deposition of 10 nm of $MoO_x$ and 100 nm of silver at $<4 \times 10^{-6}$ mbar through a 0.1 $cm^2$ pixel area shadow mask.

**Device Characterization**

*J-V* measurements of solar cells were performed in an $N_2$ filled glove box using a Keithley 2400 source meter and an Oriel Sol3A Class AAA solar simulator calibrated to 1 sun, AM1.5G, with a KG-5 silicon reference cell certified by Newport. External quantum efficiency (EQE) was characterized using an EQE system (PV measurement Inc.). Measurements were performed at zero bias by illuminating the device with monochromatic light supplied by a Xenon arc lamp in combination with a dual-grating monochromator. The number of photons incident on the sample was calculated for each wavelength by using a silicon photodiode calibrated by The National Institute of Standards and Technology (NIST).

**Light-Intensity Dependence Measurements:** Light-intensity dependence measurements were performed with PAIOS instrumentation (Fluxim) (steady-state and transient modes). Transient photo-voltage (TPV) measurements monitor the photovoltage decay upon a small optical perturbation during various constant light-intensity biases and at open-circuit bias conditions. Variable light-intensity biases lead to a range of measured $V_{oc}$ values that were used for the analysis. During the measurements a small optical perturbation (<3% of the $V_{oc}$, so that $\Delta V_{oc} << V_{oc}$) is applied. The subsequent voltage decay is then recorded to directly monitor bimolecular charge carrier recombination. The photovoltage decay kinetics of all devices follow a mono-exponential decay: $\delta V = A \exp(-t/\tau)$, where $t$ is the time and $\tau$ is the charge carrier lifetime. The light intensity is given in the maximum power of the LED source (100% ≈ 200 mW $cm^{-2}$).

**Electrochemical Impedance Spectroscopy (EIS):** was conducted using a commercially available PAIOS. The measurements were performed using conditions that had an applied bias of open circuit voltage of devices in the dark, and a frequency ranging from 3 MHz to 300 Hz.

**X-ray photoelectron spectroscopy (XPS)**: Measurements were carried out in a multi-probe Scienta Omicron Ultrahigh Vacuum system operating at a base pressure of 10 -10 mbar. The system is equipped with a monochromated $AlK^{\alpha}$ XM1000 x-ray source. The XPS electron analyzer is an Argus Cu. The XPS spectra were recorded with a constant pass energy of 20. The samples were contacted via molybdenum strips without using charge neutralization. XPS measurements were taken with the samples grounded. The measurements were conducted with the sample plane oriented normal to the analyzer direction, at 0°, 30o and 60o take-off angles, and an X-ray analyzer positioned at an angle of 80°.

**Photoelectron spectroscopy (PESA)**: The measurement were done using a Riken Keiki PESA spectrometer (Model AC-2) with a power setting of 10 nW and a power number of 0.3. Thin film samples for PESA were deposited on glass substrates.

**Films Characterizations:** The top-view scanning electron microscope (SEM) images of the films were obtained using Helios 5 UX microscope. Atomic force microscopy (AFM) was employed to characterize the topography of the films in tapping mode, using a Bruker Dimension Icon system in ambient atmosphere. XRD measurements were carried out using Bruker D2 Phaser. UV-Vis Spectra were carried out on a Cary 5000 Spectrophotometer setup in double beam mode. The optical bandgaps were estimated using direct-allowed Tauc analysis by plotting $(\alpha h\nu)^2$ as a function of photon energy $h\nu$, followed by linear extrapolation of the absorption-edge region to the energy axis.

**Density Functional Theory (DFT) calculations:** The results were obtained with the Density Functional Theory (DFT) code Vienna Ab Initio Simulation Package (VASP)[1] with plane-wave basis (and energy cutoff of 400 eV), projector augmented waves (PAW)[2] and the generalized gradient approximation (GGA) Perdew-Burke-Ernzerhof [3] exchange-correlation (xc) functional. The non-bonding van der Waals interactions were taken into account within the so called DFT-D3 scheme.[4] Structures were rendered with the software VESTA.[5]

We have performed calculations on supercells of the alpha polytype of $TeO_2$ with 72 Te atoms and 144 O atoms in the defect-free case. We first introduced a varying number of O vacancies in these supercells. The main finding in this part is that O vacancies tend to agglomerate and form thus areas of Te complexes which are embedded within the $TeO_2$ matrix. This complex formation, in turn, decreases the calculated energy band gap (as expected, given that the energy band gap of Te is significantly smaller than that of $TeO_2$). When Se atoms are inserted in these supercells, then it is energetically favorable to all Se atoms to be located in O-free areas forming Se-Te complexes. The insertion of Se modifies the calculated energy band gap (as compared to the above-mentioned cases of Te-only complexes). The main results of the DFT calculations support the role of Se doping a means to tune the structural and electronic properties of the oxide.

80 nm $TeO_x$ film

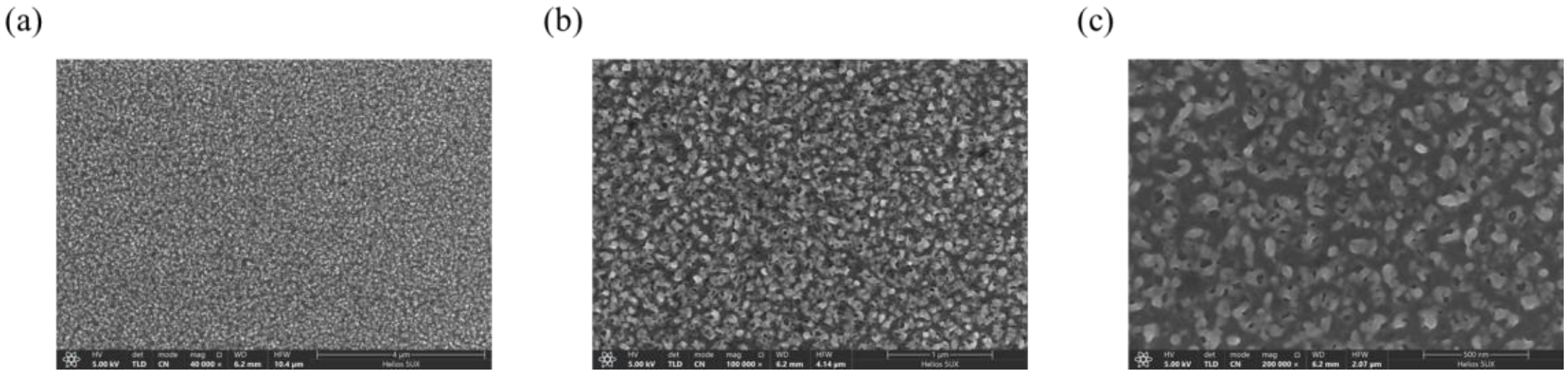

Figure S1. Top-view SEM images of the $TeO_x$ film at different magnifications. Scale bars: (a) 4 μm, (b) 1 μm, and (c) 500 nm.

80 nm $TeO_x$:Se (10 wt% Se) film, annealing for 1.5 h

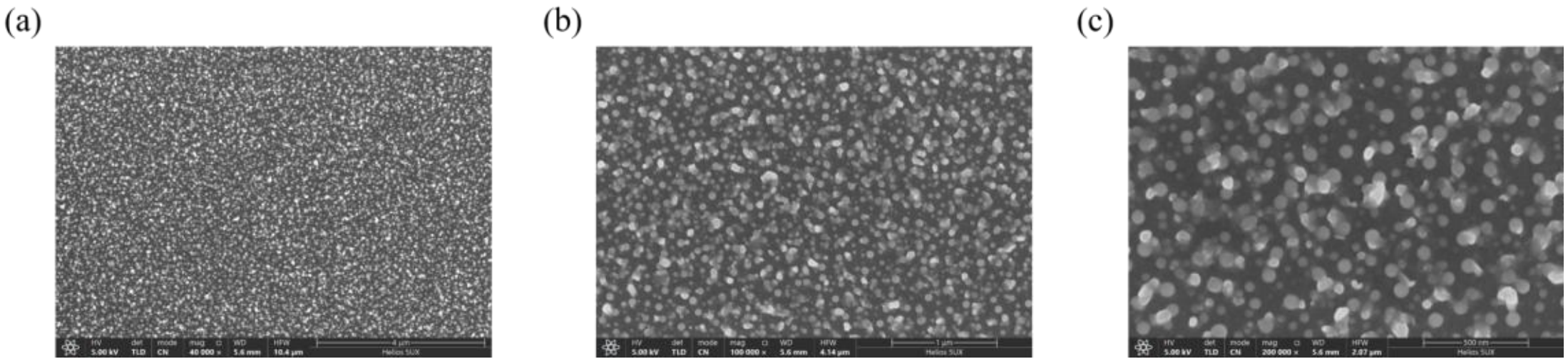

Figure S2. Top-view SEM images of the 80 nm $TeO_x$:Se (10 wt% Se) film, annealing for 1.5h at different magnifications. Scale bars: (a) 4 μm, (b) 1 μm, and (c) 500 nm.

80 nm $TeO_x$:Se (10 wt% Se) film, annealing for 9 h

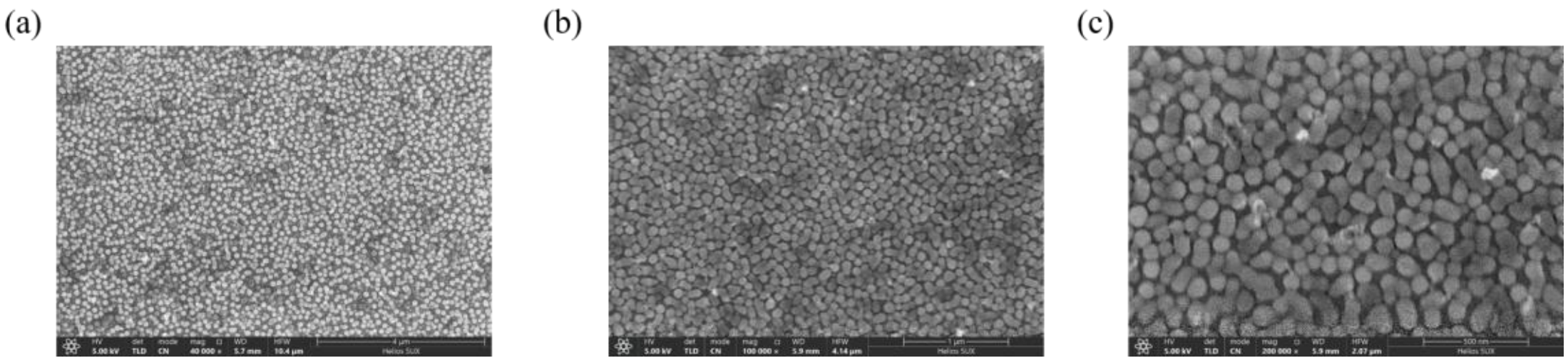

Figure S3. Top-view SEM images of the 80 nm $TeO_x$:Se (10 wt% Se) film, annealing for 9h at different magnifications. Scale bars: (a) 4 μm, (b) 1 μm, and (c) 500 nm.

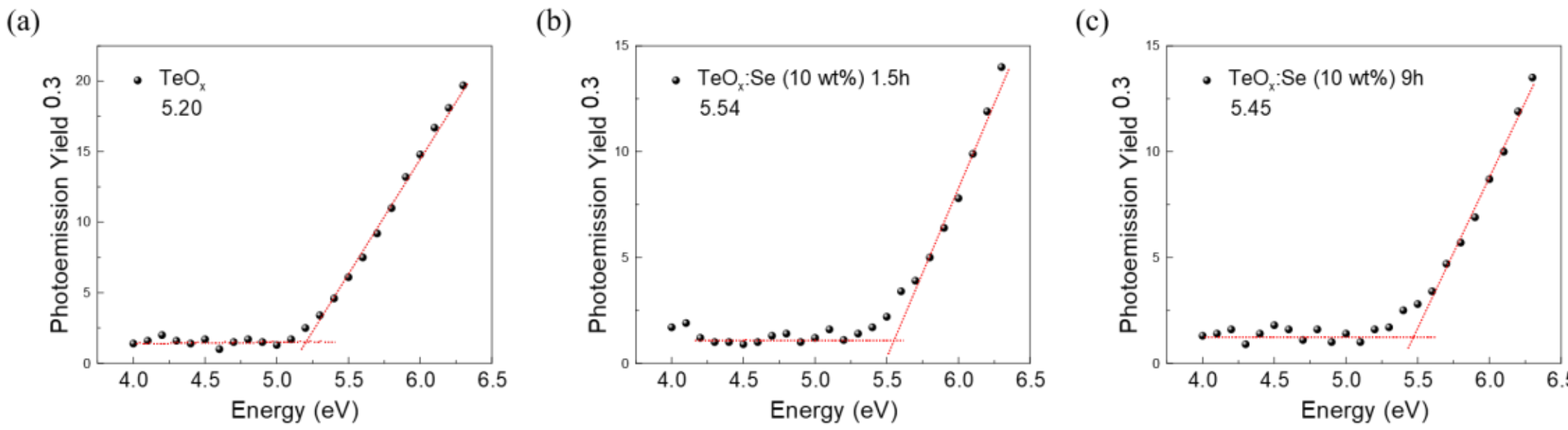

Figure S4. Determination of the VBM positions of (a) $TeO_x$, (b) $TeO_x$:Se (10 wt%) annealed for 1.5 h, and (c) $TeO_x$:Se (10 wt%) annealed for 9 h by photoelectron spectroscopy in air.

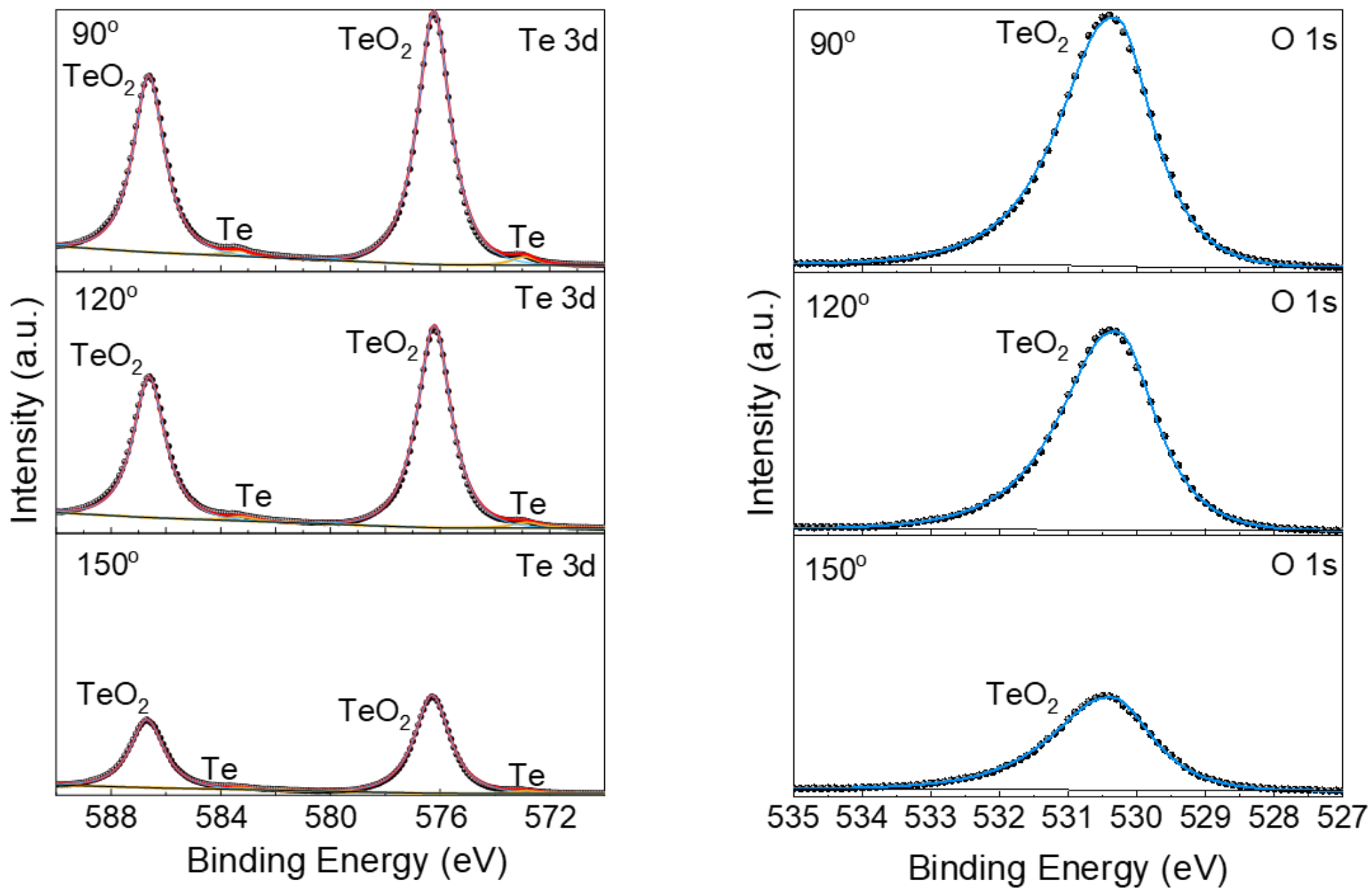

Figure S5. The XPS spectrum for $TeO_x$ samples.

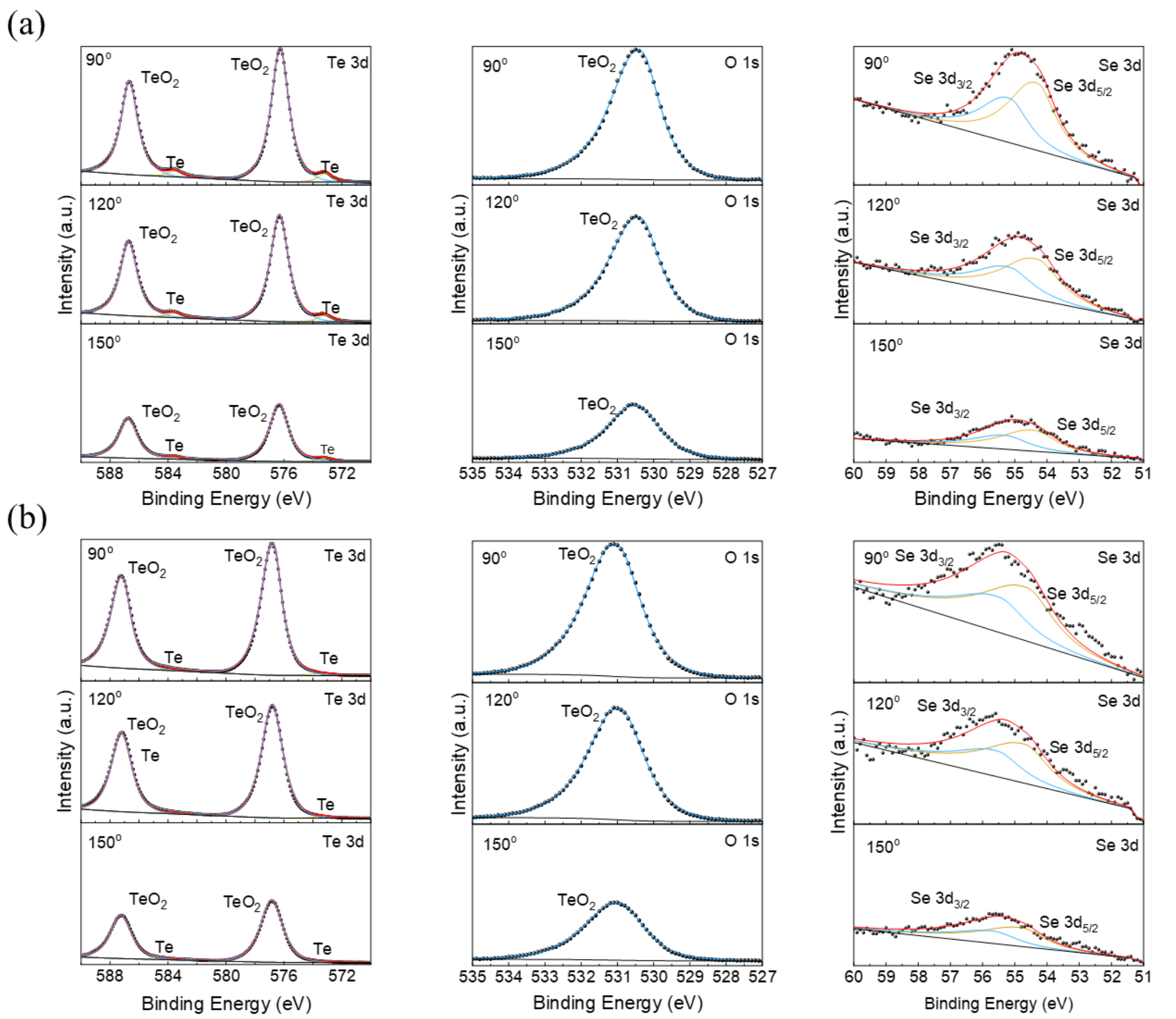

Figure S6. The XPS spectrum for $TeO_x$:Se (10 wt% Se, 1.5 h) sample and $TeO_x$:Se (10 wt% Se, 9 h) sample.

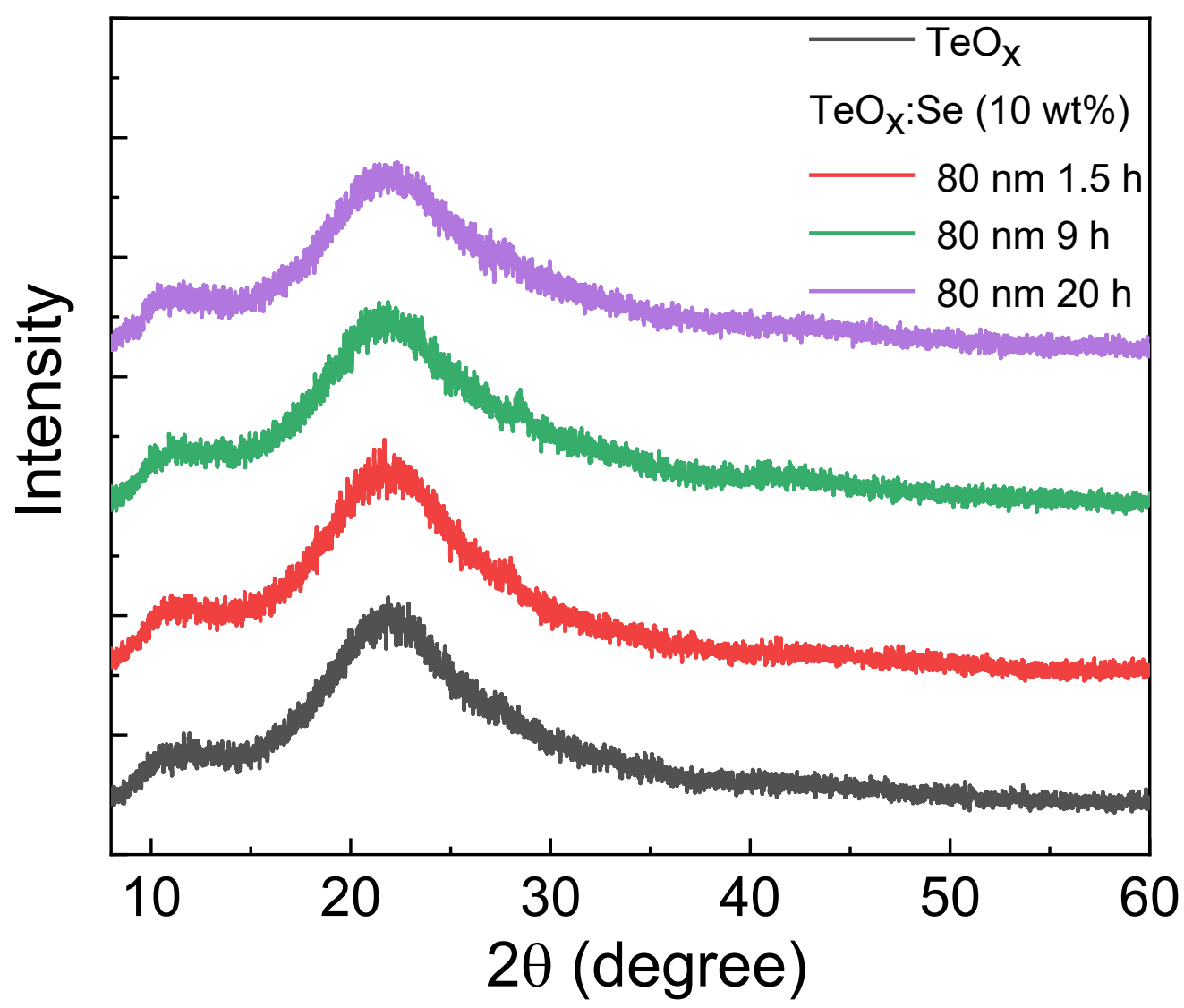

Figure S7. XRD image of 80 nm various samples.

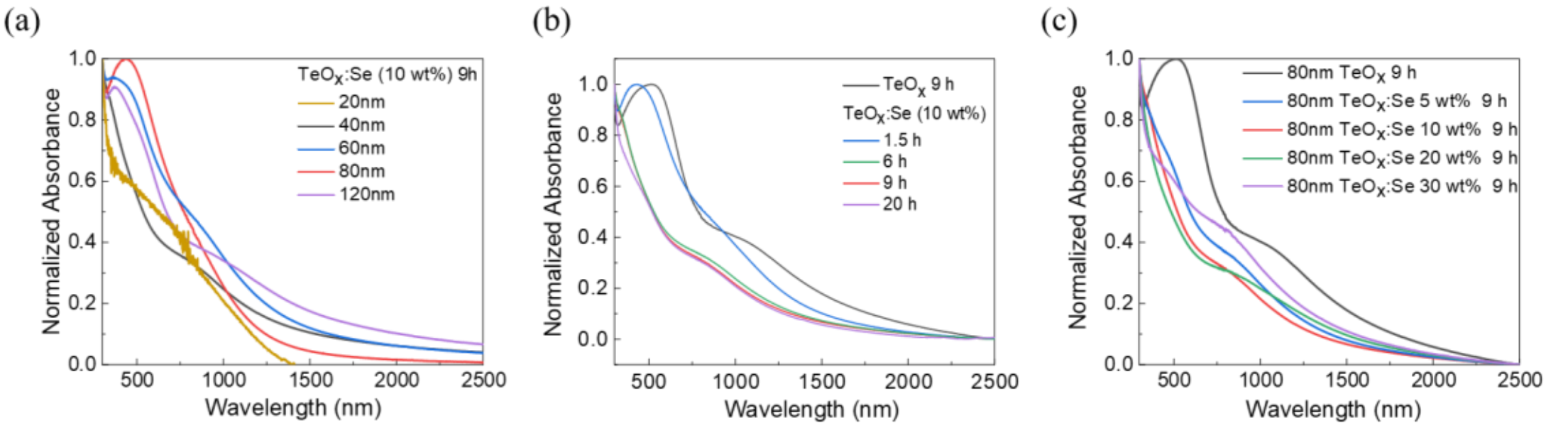

Figure S8. The absorption spectra of (a) different thickness $TeO_x$:Se (10 wt% Se) annealing for 9 h, (b) 80 nm $TeO_x$:Se (10 wt% Se) annealing for different time, (c) different Se concentration $TeO_x$:Se samples annealing for 9 h

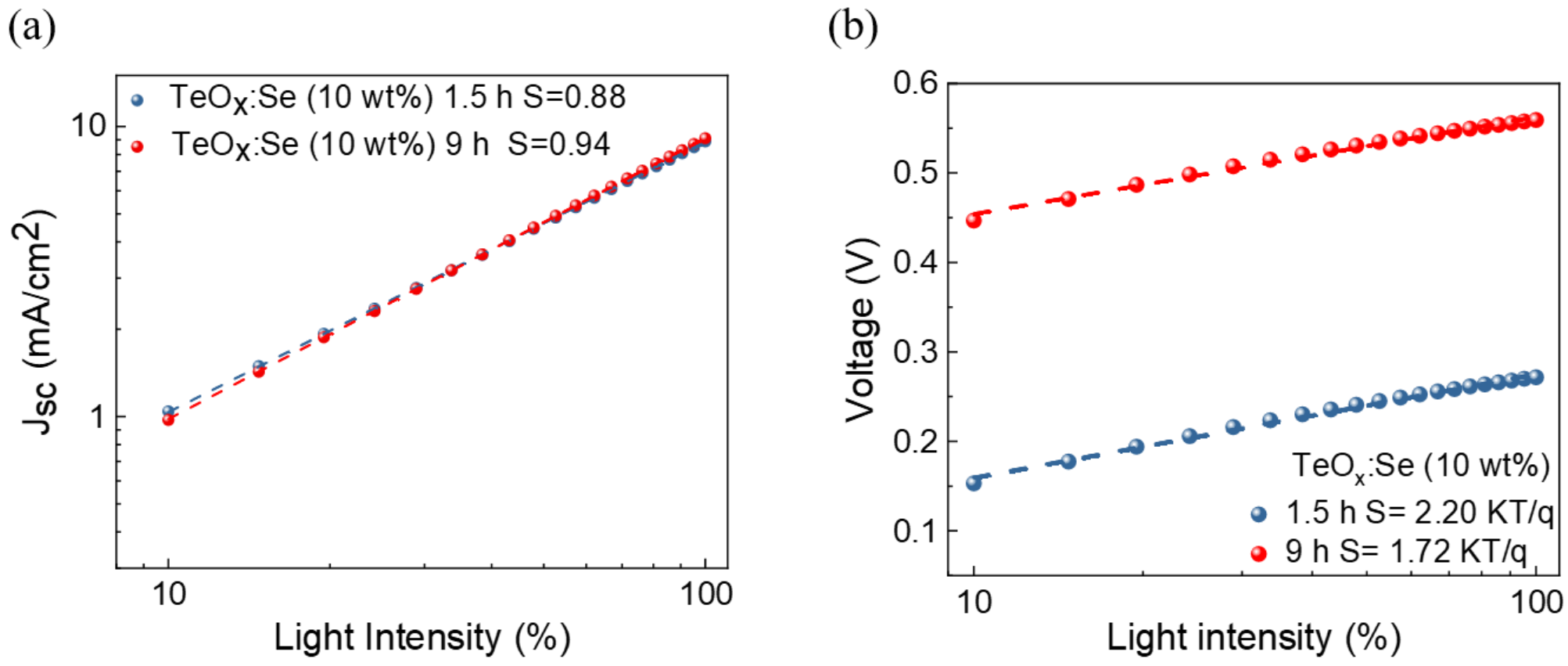

Figure S9. (a) $J_{SC}$ of the $TeO_x$:Se (10 wt% Se) annealing at 1.5 h and 9 h dependent on light intensity, (b) *Voc* of the $TeO_x$:Se (10 wt% Se) annealing at 1.5 h and 9 h dependent on light intensity.

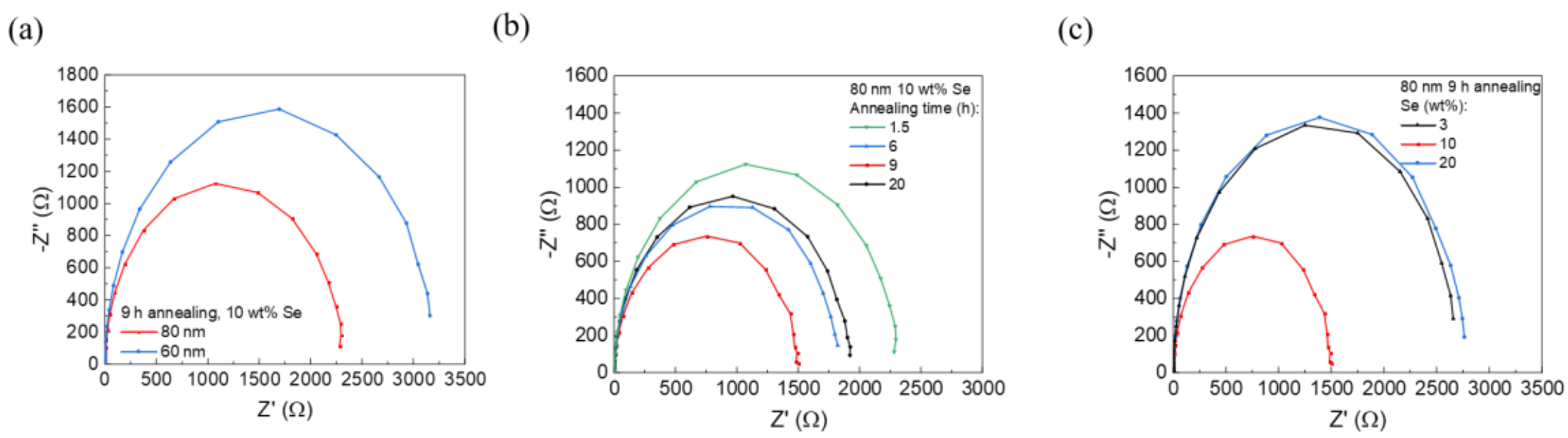

Figure S10. Electrical impedance spectroscopy (EIS) measurement for $TeO_x$:Se at (a) different thickness, (b) different annealing time, (c) Se concentration.

| TeOx | Name | Position | FWHM | Raw Area | % At Conc | $O/Te^{4+}$ |
|---|---|---|---|---|---|---|
| **$90^o$** | **O** | 530.31 | 1.26 | 41359.09 | 65.7 | 2.0 |
| | **$Te^{4+}$** | 576.22 | 1.37 | 170165.68 | 33.5 | |
| | **Te** | 572.95 | 1.08 | 4168.96 | 0.8 | |
| **$120^o$** | **O** | 530.28 | 1.27 | 33458.09 | 65.8 | 2.0 |
| | **$Te^{4+}$** | 576.20 | 1.39 | 137438.64 | 33.5 | |
| | **Te** | 572.90 | 1.08 | 2980.93 | 0.7 | |
| **$150^o$** | **O** | 530.35 | 1.35 | 17336.46 | 66.3 | 2.0 |
| | **$Te^{4+}$** | 576.28 | 1.45 | 69760.11 | 33.1 | |
| | **Te** | 572.89 | 1.07 | 1345.28 | 0.6 | |

Table S1. The XPS information for $TeO_x$ 9h samples.

| $TeO_x$:Se | Name | Position | FWHM | Raw Area | % At Conc | $O/Te^{4+}$ |
|---|---|---|---|---|---|---|
| **$90^o$** | **O** | 530.37 | 1.29 | 34827.46 | 61.9 | 2.0 |
| | **$Te^{4+}$** | 576.26 | 1.39 | 141769.95 | 31.2 | |
| | **Te** | 573.15 | 1.19 | 7756.58 | 1.7 | |
| | **Se** | 55.10 | 1.62 | 1890.71 | 5.2 | |
| **$120^o$** | **O** | 530.39 | 1.31 | 28374.00 | 62.2 | 2.0 |
| | **$Te^{4+}$** | 576.30 | 1.42 | 115367.61 | 31.3 | |
| | **Te** | 573.19 | 1.17 | 5890.80 | 1.6 | |
| | **Se** | 55 | 1.65 | 1446.07 | 4.9 | |
| **$150^o$** | **O** | 530.41 | 1.36 | 16093.50 | 61.9 | 2.0 |
| | **$Te^{4+}$** | 576.33 | 1.49 | 65230.25 | 31.1 | |
| | **Te** | 573.18 | 1.13 | 2822.63 | 1.4 | |
| | **Se** | 55.10 | 1.66 | 961.92 | 5.7 | |

Table S2. The XPS information for $TeO_x$:Se (10 wt% Se, 1.5 h) samples.

| $TeO_x$:Se | Name | Position | FWHM | Raw Area | % At Conc | $O/Te^{4+}$ |
|---|---|---|---|---|---|---|
| **$90^o$** | **O** | 531.02 | 1.55 | 36552.39 | 64.1 | 2.1 |
| | **$Te^{4+}$** | 576.86 | 1.65 | 143344.61 | 31.1 | |
| | **Te** | 573.46 | 1.39 | 1879.37 | 0.4 | |
| | **Se** | 55.40 | 2.36 | 1633.98 | 4.4 | |
| **120°** | **O** | 530.92 | 1.56 | 31061.58 | 63.9 | 2.0 |
| | **$Te^{4+}$** | 576.80 | 1.67 | 123065.14 | 31.4 | |
| | **Te** | 573.45 | 1.38 | 1388.02 | 0.4 | |
| | **Se** | 55.50 | 2.32 | 1383.49 | 4.4 | |
| **150°** | **O** | 530.96 | 1.62 | 17501.35 | 63.1 | 2.0 |
| | **$Te^{4+}$** | 576.83 | 1.74 | 70316.13 | 31.4 | |
| | **Te** | 573.20 | 1.44 | 819.59 | 0.4 | |
| | **Se** | 55.40 | 2.34 | 920.62 | 5.1 | |

Table S3. The XPS information for $TeO_x$:Se (10 wt% Se, 9 h) samples.

| **Sample** | **PESA-derived VBM (eV)** | **Optical bandgap (eV)** | **Estimated CBM (eV)** |
|---|---|---|---|
| $TeO_x$ | -5.20 | 0.86 | -4.34 |
| $TeO_x$:Se (10 wt% Se, 9 h) | -5.45 | 0.98 | -4.47 |
| $TeO_x$:Se (10 wt% Se, 1.5 h) | -5.54 | 1.09 | -4.45 |

Table S4. The PESA-derived VBM, optical bandgap and estimated CBM for different Samples.

| Annealing time (h) | Thickness (nm) | $J_{sc}$(mA/cm$^2$) | $V_{oc}$(V) | FF | PCE (%) |
|---|---|---|---|---|---|
| 1.5 | 20 | x | | | |
| | 40 | x | | | |
| | 60 | 4.08 | 0.19 | 0.25 | 0.19 |
| | 80 | 5.69 | 0.26 | 0.35 | 0.52 |
| | 120 | x | | | |

Table S5. Summary of photovoltaic parameters of different thickness $TeO_x$:Se (10 wt% Se) samples annealing for 1.5 h, measured under one-sun illumination. X means device

doesn't work.

| Thickness (nm) | Annealing time (h) | $J_{sc}$ (mA/cm$^2$) | $V_{oc}$(V) | FF | PCE(%) |
|---|---|---|---|---|---|
| 80 | 1.5 | 5.69 | 0.26 | 0.35 | 0.52 |
| | 6 | 5.47 | 0.34 | 0.33 | 0.61 |
| | 9 | 8.50 | 0.50 | 0.44 | 1.87 |
| | 20 | 5.95 | 0.41 | 0.34 | 0.83 |

Table S6. Summary of photovoltaic parameters of 80 nm $TeO_x$:Se (10 wt% Se) samples annealing at different time, measured under one-sun illumination.

| Annealing time (h) | Thickness (nm) | Se (%) | $J_{sc}$ (mA/cm$^2$) | $V_{oc}$ (V) | FF | PCE (%) |
|---|---|---|---|---|---|---|
| 9 | 80 | 0 | x | | | |
| | | 3 | 6.5 | 0.20 | 0.24 | 0.31 |
| | | 10 | 8.5 | 0.5 | 0.44 | 1.87 |
| | | 20 | 7.51 | 0.44 | 0.25 | 0.83 |

Table S7. Summary of photovoltaic parameters of 80 nm, 9 h annealing time $TeO_x$:Se samples for different Se concentrations, measured under one-sun illumination.